\documentclass[11pt]{article}
\usepackage{color}
\usepackage{amssymb}
\usepackage{amsmath}
\usepackage{multirow}
\usepackage{epsfig}
\usepackage{amscd}

\def\Journal#1#2#3#4{{#1} {\bf #2}, #3 #4}

\def\etal{{\it et al.}}

\def\NPB{{\em Nucl. Phys.} B}
\def\NPS{\em Nucl.Phys.Proc.Suppl.}

\def\PRD{{\em Phys. Rev.} D}

\def\be{\begin{equation}}
\def\ee{\end{equation}}
\def\bea{\begin{eqnarray}}
\def\eea{\end{eqnarray}}

\def\mg{\mathsf g}

\def\um{\mathcal U}

\begin{document}
{\Large \bf Dark energy condensate and vacuum energy}\\

{\bf Houri Ziaeepour}, {\it Max Planck Institut f\"ur Extraterrestrische Physik (MPE), 
Giessenbachstra$\mathbf{\beta}$e 1, 85748 Garching, Germany.}\\


\section* {\normalsize Abstract}
Many candidate models for dark energy are based on the existence of a classical scalar field. 
In the context of Quantum Field Theory (QFT), we briefly discus the condensation of such a field 
from a light quantum scalar field produced by gradual decay of a heavy particle during 
cosmological time. We obtain the necessary conditions for survival of the condensate in an 
expanding universe and show that this process is directly related to quantum nature of the 
field which preserves the coherence of the condensate at cosmological distances. We also 
suggest a new interpretation of {\it vacuum energy} in QFT in curved space time which can 
potentially solve the puzzle of huge deviation of what is considered to be the vacuum energy 
from observations of dark energy.

\section {\normalsize Introduction} \label{sec:intro}
Many alternatives to a cosmological constant have been proposed to explain the accelerating 
expansion of the Universe. They can be divided to two main groups: modified gravity models and 
models in which a field - usually a scalar but also in some cases a vector field - is 
responsible for what is called dark energy. This limited contribution does not allow 
to go to the details of each category, therefore we only concentrate on the models based on a 
scalar field, generally called {\it quintessence}~\cite{quinmodel0,quinmodel1}. In simplest 
version of such models the quintessence field $\phi$ is a very light scalar with a 
self-interaction potential and no interaction with other components of the Universe. Under 
special conditions~\cite{trackingcond} the dynamics of the field at late times become 
independent of its initial value and the field approaches to what is called a tracking solution. 
In this case it varies very slowly with time and its equation of state, defined as:
\be
w = \frac {P}{\rho} = \frac {\frac{1}{2}\dot{\phi}^2 - V (\phi)}{\frac{1}{2}\dot{\phi}^2 + 
V (\phi)} \label{quinw}
\ee
approaches $w \gtrsim -1$. Two types of potentials have tracking solutions: $V (\phi) = 
e^{-\alpha \phi}$, ${\phi}^{-n}$, and polynomials including such terms. In the context of QFT these 
potentials are non-renormalizable, thus are assumed to be effective potentials.

A simple quintessence model suffers from various short comings. One of these problems is the 
very small mass of $\phi$ which must be $m_\phi \sim 10^{-32}$ eV $\sim H_0$. More importantly, 
this model cannot explain what is called {\it the coincidence problem}, i.e. why dark energy 
becomes dominant only after galaxy formation. This means that the density fraction of dark 
energy at the time of matter formation - presumably after inflation and during reheating - had 
to be $\sim 10^{42}$ times smaller than matter density. The only natural way to explain such an 
extreme fine tuning is to consider an interaction between dark energy and other components, 
notable with dark matter~\cite{quinint,houriustate,houridmquin0}. Moreover, a simple 
quintessence model is limited to $w > -1$, but for the time being many observations prefer 
$w \lesssim -1$, although due to measurement errors one cannot yet have a definitive conclusion 
about the sign of $w+1$. Interacting dark energy models can explain $w < -1$ without violation 
of null energy principle because it has been shown~\cite{houriustate,quindmint} that when the 
interaction is ignored, the effective equation of state $w_{eff} < -1$ when the real $w \geq -1$. 
As for the particle physics view, no particle is isolated and every species has some 
non-gravitational interaction with other particles.

The study of quintessence models is usually concentrated on the evolution of a classical scalar 
field without any concern about how such a field can be formed from a quantum field, specially 
in an expanding universe. In fact considering very small mass of a quintessence field and its 
very weak interaction to itself and to other particles, one expects that at their production - 
during inflation, reheating and/or later in the history of the Universe - they simply behave 
as relativistic particles and have an equation of state $w \sim 1/3$ which is very different 
from dark energy $w \approx -1$. Nonetheless, we also know that bosonic particles/fields can 
condensate and 
form a classical scalar field. We know this process from condense matter where Cooper pairs 
creates a non-zero expectation value - a condensate - at macroscopic scales, breaks the 
$U(1)$ symmetry, generates an effective mass for photons, and leads to phenomena such as 
superconductivity and super-fluidity. The Higgs field - if it exists - has a similar property 
but at microscopic scales. The formation of Higgs condensate at electroweak energy scale breaks 
$SU(2) \times U(1)$ and generates mass for leptons and quarks. The formation of a  condensate 
has been studied, see e.g.~\cite{composhiggs} for some Higgs models as well as for 
inflaton~\cite{infcondens}. These studies show that the problem of condensate formation and 
evolution is quite involved. In the case of dark energy it is even more complicated because 
one has to take into account the geometry of the expanding Universe and the evolution of other 
species, specially in the context of interacting quintessence models. The condensation issue is 
also more important because in contrast to Higgs and inflation, dark energy condensate must be 
very uniform and homogeneous both spatially and during cosmic time. These properties cannot be 
obtained trivially and should strongly constrain quintessence models. 

In this proceeding we briefly review the technique, issues, and results obtained recently for 
a simple and generic interacting quintessence model~\cite{houricondens}. We also describe an 
idea about a modified definition of vacuum energy which can solve the enormous deviation of 
the value obtained from usual definition in QFT.

\section{\normalsize Dark energy from decay of dark matter} \label{sec:dmquin}
\begin{figure}
\begin{center}
\begin {tabular}{llll}
\includegraphics[height=4cm,angle=-90]{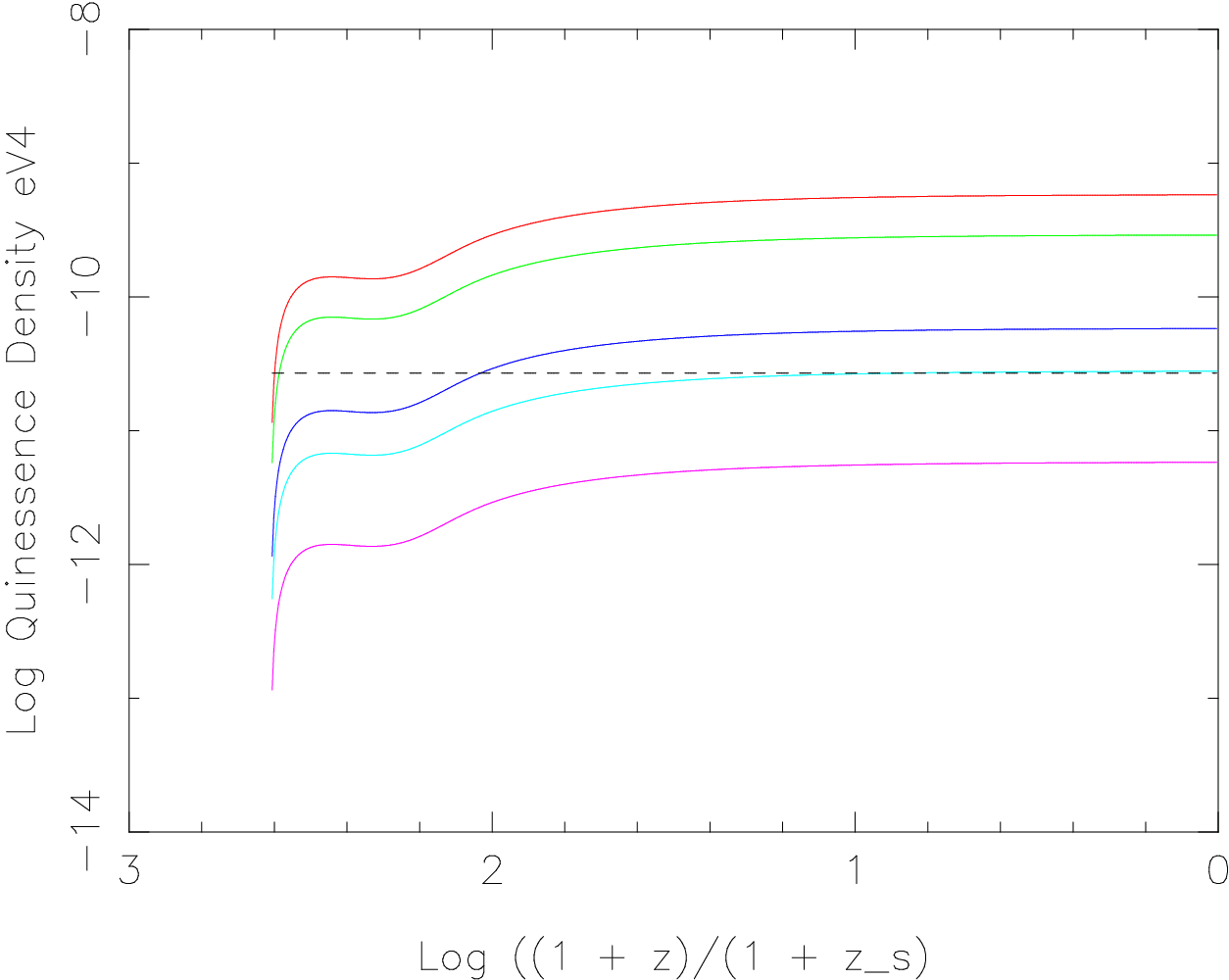}
\includegraphics[height=4cm,angle=-90]{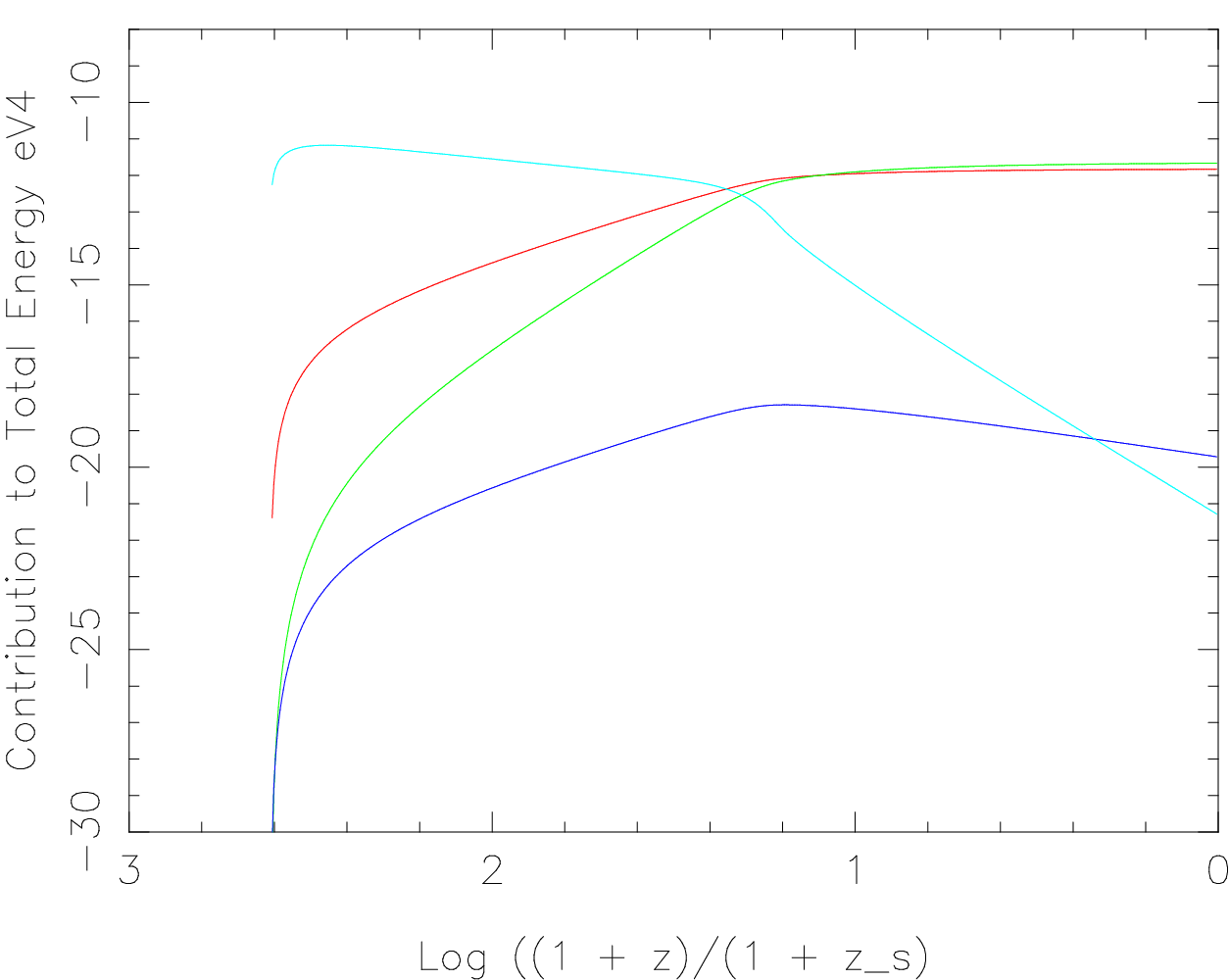}
\includegraphics[height=4cm,angle=-90]{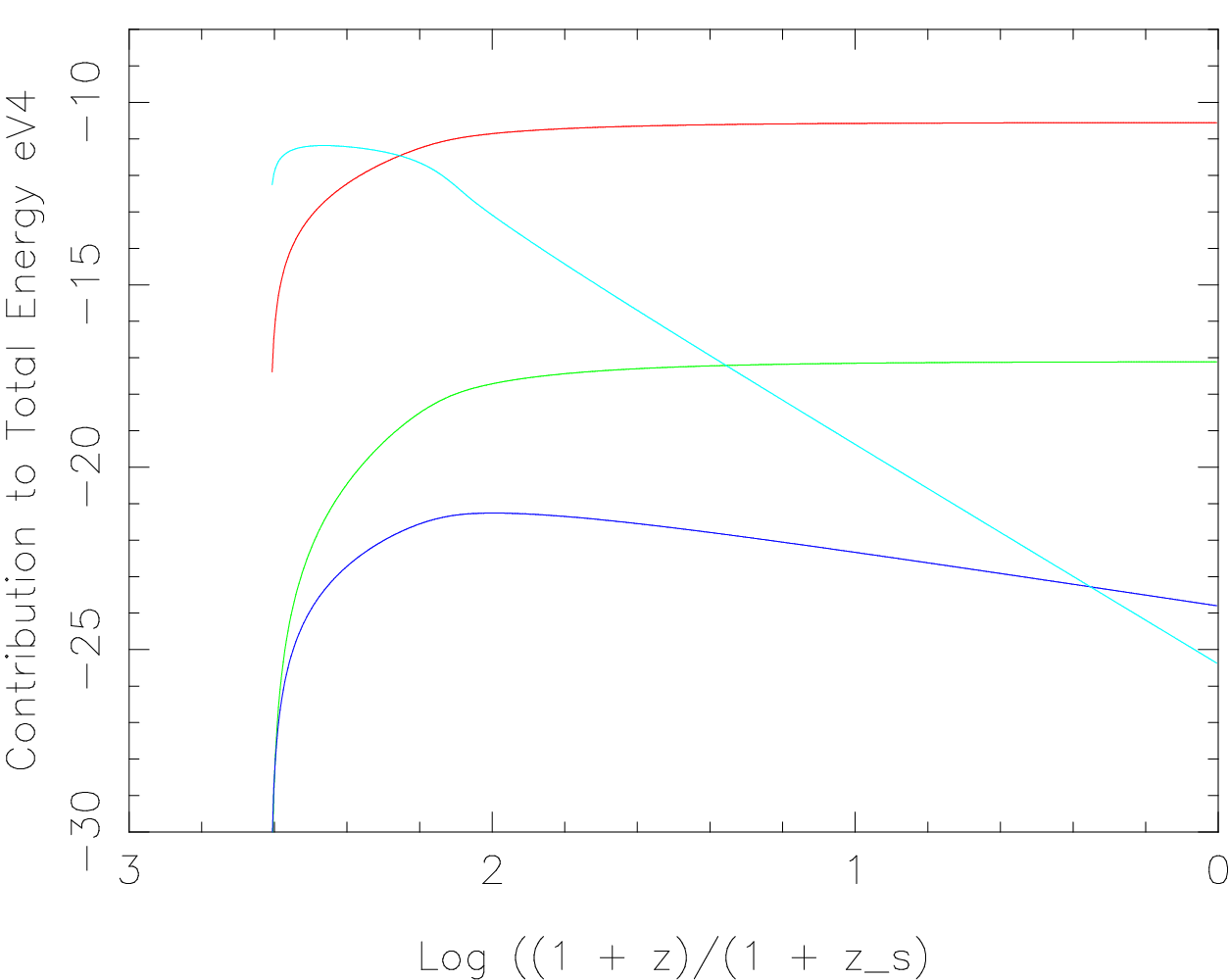}
\includegraphics[height=4cm,angle=-90]{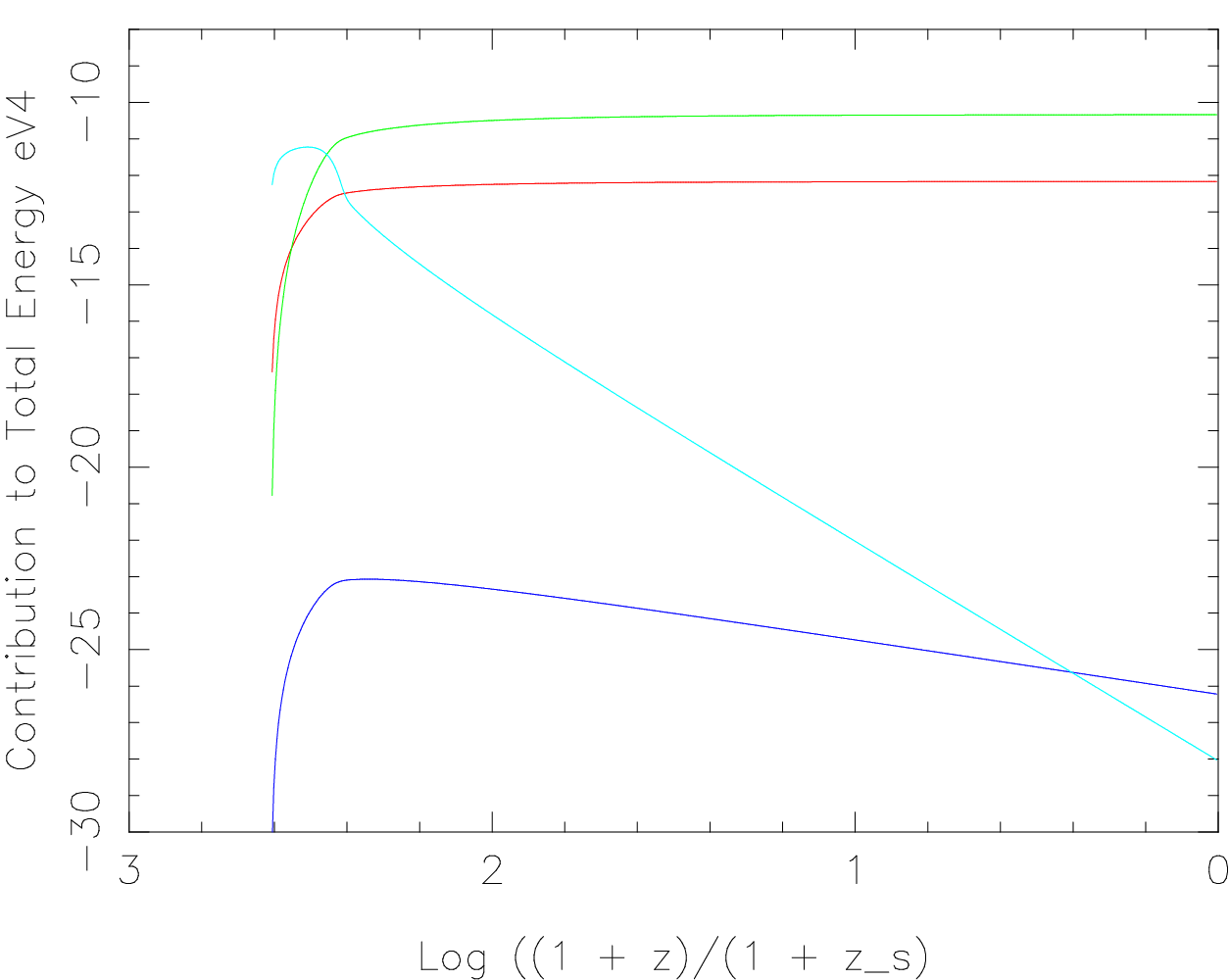}
\end{tabular}
\caption {From left to right: 1) Density of dark energy for various branching ration to the 
quintessence field ${\Gamma}_0 \equiv {\Gamma}_\phi/\Gamma = 10^{-16}$ (magenta), $5 {\Gamma}_0$ 
(cyan), $10 {\Gamma}_0$ (blue), $50 {\Gamma}_0$ (green), $100 {\Gamma}_0$ (red). 
Dash line is the observed value of the dark energy. $m_\phi = 10^{-6} eV$, self-coupling 
$\lambda = 10^{-20}$. 2,3,4) Evolution of the contribution to the total energy density of $\phi$ 
for ${\Gamma}_0 = 10^{-16}$ and 2) $m_\phi = 10^{-8} eV$ and $\lambda = 10^{-20}$; 
3) $m_\phi = 10^{-6} eV$ and $\lambda = 10^{-20}$; 4) $m_\phi = 10^{-6} eV$ and $\lambda = 10^{-10}$. 
Curves are: mass (red), self-interaction (green), kinetic energy (cyan) and interaction with 
DM (blue). \label {fig:dmquin}}
\end{center}
\end{figure}

In the context of inflation-reheating models, all constituents of the Universe were produced 
either during reheating from the decay of inflaton or a curvaton field, or later on from the 
decay of other species. In~\cite{houridmquin0} we have studied the decay of a massive long life 
metastable dark matter with a small branching ratio to a light scalar field. A classical 
treatment of such a model show that the energy density of the light scalar field from very 
early times is roughly constant despite the expansion of the Universe, i.e. it behaves very 
similar to a cosmological constant, see Fig. \ref{fig:dmquin}. In contrast to many quintessence 
models, in this model the self-interaction potential is a simple $\phi^4$ polynomial. The 
scalar field has a $w \sim -1$ for a large range of parameters and is not very sensitive to 
self-interaction potential because it is mainly the interaction/decay term that control its 
evolution during cosmic time. Indeed, such a setup has an internal feedback - if the density of 
dark energy increases, expansion rate increases, reduces the density of dark matter and 
thereby the rate of production of scalar field from decay of dark matter, thus the density of 
dark energy decreases. For a metastable dark matter, this stability can last for very long time. 
In addition, there would be no big ripe in the future because when a large fraction of dark 
matter decays, the stability of the system breaks, the energy density of dark energy decreases 
and the Universe becomes matter or radiation dominated, thus the accelerating expansion rate 
slow downs. Such a 
model can easily explain the observed slightly negative value of 
$w+1$~\cite{houriustate,quindmint}. Therefore, for studying the condensation of quintessence 
field we consider this model.

\section{\normalsize Condensation of an interacting dark energy} \label{sec:condens}
In condense matter a condensate is defined as a system in which the majority of particles are 
in their ground state. In quantum field theory when there is no conserved quantum number, such 
as in the case of a single scalar field without internal symmetry and with self interaction, 
a system is not usually in an eigen state of the number operator. Therefore a condensate which 
behaves classically is defined as a state in which number operator has a large expectation 
value - large occupation number - equivalent to a classical system with a large number of 
particles - in the minimum of their potential energy. Mathematically a condensate state 
$|\Psi\rangle$ is defined as:
\be
\langle \Psi|\phi |\Psi \rangle \equiv \varphi \neq 0 \label{conddef}
\ee
It is easy to verify that in contrast to Bose-Einstein condensate in quantum mechanics, for a 
quantum system containing only free or weakly interacting - perturbative - fields with finite 
number of particles $\varphi$ is zero. It is possible to construct states which satisfies 
equation (\ref{conddef}) using an expansion to coherent states. A special case is suggested 
by~\cite{condwave} and a more general state that we call {\it multi-condensate} is obtained 
in~\cite{houricondens}. For a real scalar field it has the following expression:
\bea
&& |\Psi_{GC}\rangle \equiv \sum_k A_k e^{C_k a_k^{\dagger}} |0\rangle = \sum_k A_k 
\sum_{i=0}^{N \rightarrow \infty} \frac {C_k^i}{i!}(a_k^{\dagger})^i |0\rangle 
\label{condwaveg} \\
&& \chi (x,\eta) \equiv a (\eta) \langle \Psi_{GC}|\Phi|\Psi_{GC}\rangle = \sum_k  
C_k \um_k (x) + C^*_k \um_k^* (x) \quad \Longrightarrow \quad C_k = \frac{\um_k (x) + 
\um_k^* (x)}{\chi (x)} \label{condexpg}
\eea
where operators $a_k$ and $a_k^{\dagger}$ are respectively annihilation and creation operators 
with $[a_k,a_k^{\dagger}] = 1$. Coefficients $A_k$ and $C_k^i$ are arbitrary, but can depend on 
the spacetime coordinate because creation and annihilation operators in a curved space depend 
on the coordinates. $\um_k$ is a solution of the free Green's function of $\phi$.

To study the formation and evolution of such a state we consider a toy model for the Universe 
after reheating. Inspired by the classical model explained in the previous section, we assume 
a heavy particle - for the sake of simplicity a scalar $X$ - presumably the dark matter, that 
decays to a light scalar field $\phi$ and some other particles that we collectively call $A$. 
In general $\phi$ can have a self-interaction considered here to be a simple power-law with 
positive exponent $V(\phi) = \phi^n,~ n > 0$. For $X$ we consider only a mass term and 
no self-interaction. The collective field $A$ can have a self-interaction too. The field $\phi$ 
can be decomposed to $\phi = \Phi + I \varphi$ with $\langle \Psi|\Phi |\Psi \rangle = 0$, $I$ 
the unit operator, and $\varphi (x)$ a classical scalar field (a $\mathcal {C}$-number). 
According to this definition $\langle \Psi|\phi |\Psi \rangle = \varphi (x)$. After inserting 
this decomposition to the Lagrangian, the dynamic equation of the classical field can be obtained 
from variation principle:
\be
\frac{1}{\sqrt{-g}}{\partial}_{\mu}(\sqrt{-g} g^{\mu\nu}{\partial}_{\nu} \varphi) + 
m_{\Phi}^2 \varphi + \frac{\lambda}{n}\sum_{i=0}^{n-1} (i+1) \binom{n}{i+1}{\varphi}^i
\langle{\phi}^{n-i-1}\rangle - \mg \langle XA\rangle = 0 \label{dynphi}
\ee
A proper solution of this apparently simple equation needs a complete solution of Boltzmann or 
Kadanov-Baym equations - if we want to consider full nonequilibrium quantum field. This is a 
very complex problem specially in an expanding universe with a curved spacetime, and needs 
numerical solution of all the coupled equations. Therefore, rather than considering the full 
formulation, we assume that the evolution of other components affects $\varphi$ only through 
the expansion rate of the Universe $a(t)$ in a FLRW cosmology with metric:
\be
ds^2 = dt^2 - a^2(t){\delta}_{ij}dx^idx^j = a^2(\eta) (d{\eta}^2 - {\delta}_{ij}dx^idx^j), 
\quad dt \equiv a d{\eta} \label{metric}
\ee
This approximation is applicable to radiation domination and matter domination epochs, but not 
to redshifts $z \lesssim 1$ when the dark energy becomes dominant. Although the decoupling of 
$a(t)$ simplifies the problem, we have yet to consider a state, or in classical limit a 
distribution, for other fields to determine the expectation values in equation (\ref{dynphi}). 
Fowllowinf cosmological observations, we assume a thermal distribution for other components.
Evidently this is a valid 
assumption only at redshifts much smaller than reheating epoch, but due to the long lifetime 
of $X$ particles only a negligible fraction of them decay earlier and their impact on the later 
state of matter should be small.

We solve equation (\ref{dynphi}) separately for radiation and matter domination epochs because 
they have very different evolution equations. To determine expectation values, we use Schwinger 
closed time path integral method, but we only consider tree level diagrams and only need to 
determine free propagators. Considering the very weak coupling of $\phi$, this is a good 
approximation, up to the precision we need here. The Green's function of $\phi$ is coupled to 
the classical field $\varphi$, but considering the smallness of the self-coupling $\lambda$, 
we first determine the solution without taking into account the coupling term, then we use WKB 
approximation to obtain a more precise solution. Finally, we use $\alpha$-vacuum 
as the initial condition for the Green's function.

\begin{figure}
\begin{center}
\begin {tabular}{cc}
\includegraphics[height=4cm,angle=-90]{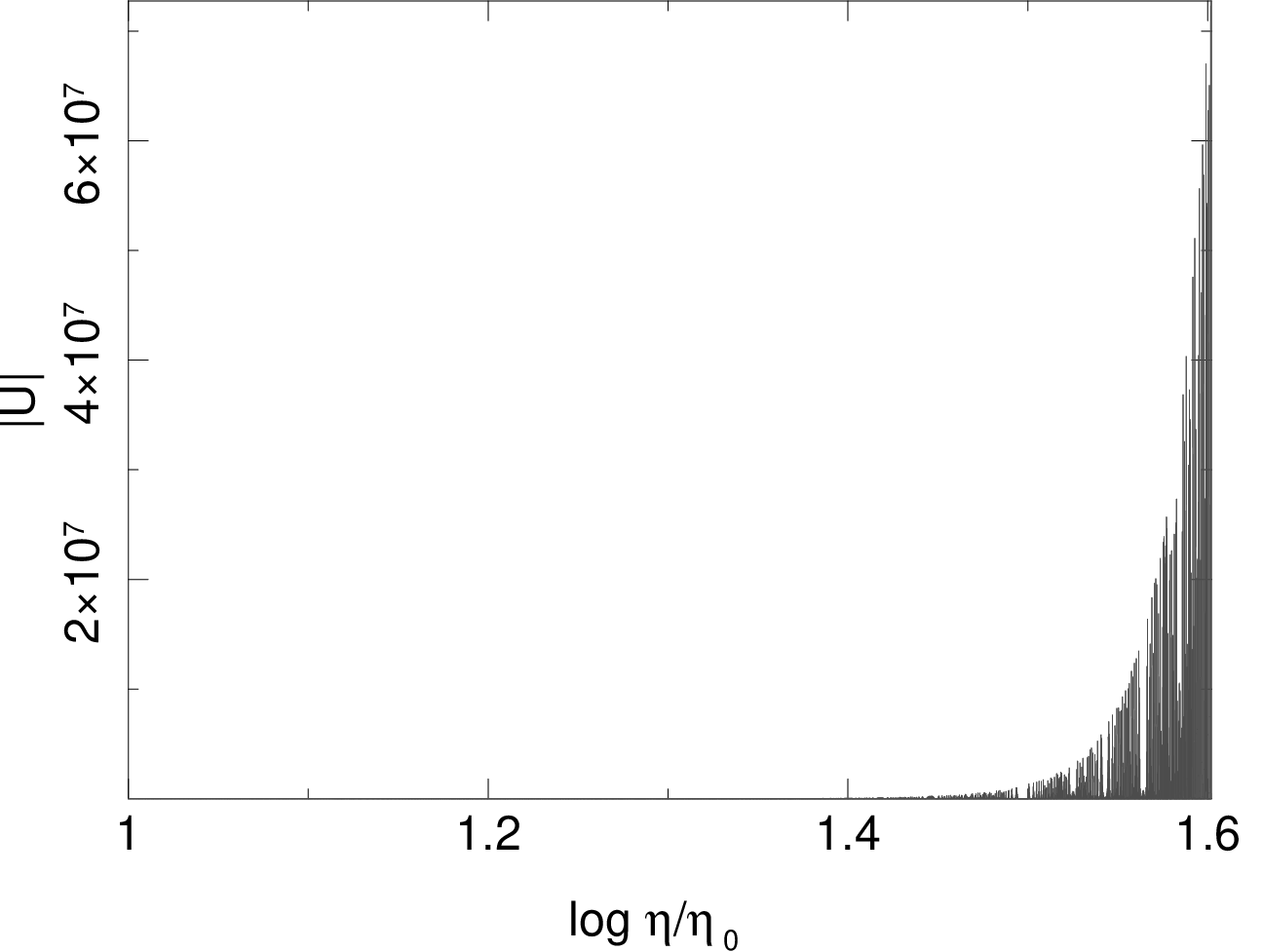} &
\includegraphics[height=4cm,angle=-90]{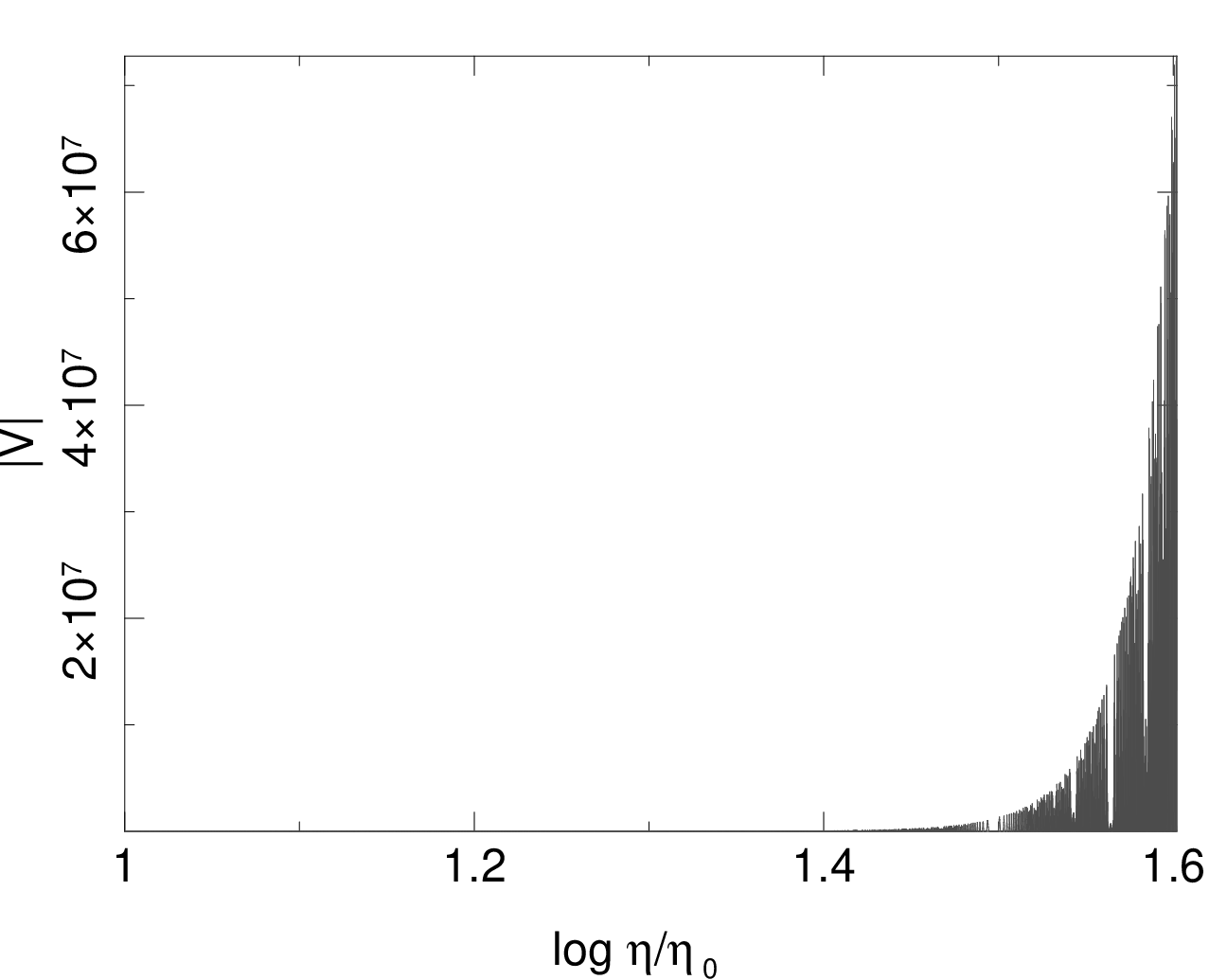} 
\end {tabular}
\end {center}
\caption{an example of absolute value of independent solutions of evolution function of $\varphi$
Note that although there are resonant jumps in the solution, due to the complexity of the 
interaction terms they are not regular like in preheating case.
\label{fig:uv}}
\end{figure}

During radiation domination epoch the Green's function equation without $\varphi$ term has an 
exact solution. The evolution equation for $\varphi$ also has the same form when interactions, 
including expectation values, are neglected. Their effect can be added through a WKB 
approximation. We consider an initial value $\varphi (t_0) = 0$ for the condensate. Finally, we 
obtain two independent solutions of the evolution equation which are plotted in 
Fig. \ref{fig:uv}. As this figure shows, the amplitude of the condensate has an exponential 
growth, similar to what happens during preheating and resonant decay of inflaton to other 
fields. This is not a surprise because $X$ and $\phi$ have a relation analogue 
to inflaton and matter fields, and have very similar evolution equations. Evidently, the 
exponential growth of the amplitude cannot continue forever, and backreactions due to 
nonlinearities in the evolution equation stop the growth rate. In particular, the interaction 
between the condensate component and {\it free $\phi$} particles through nonlinear 
self-interaction terms in (\ref{dynphi}) has the tendency to free particles from condensate, 
in another word when the density of condensate grows, it begins to {\it evaporate}. Due to 
their tiny mass, free particles are relativistic and with the expansion of the Universe they 
become diluted very quickly. On the other hand, if a large number of them {\it evaporate}, 
their energy loss during the expansion of the Universe increases the probability of joining 
the condensate again. Therefore, as long as the expansion of the Universe is not very quick, 
this process is self-regulatory. It can be shown that the amplitude of modes decreases 
very rapidly with increasing $|k|$, i.e. for small distance scales. This is consistent with 
the lack of significant spatial fluctuation in dark energy density.

In the same way one can solve the Green's function and evolution equations during matter 
domination epoch. However, even when the interactions are ignored, these equations have 
a known analytical solution only if $m_\phi = 0$ or $k = 0$. Because we are specially 
interested in the modes with $|k| \rightarrow 0$ (Similar to radiation domination epoch, it is 
possible to show that the amplitude of modes for large $|k|$'s decreases quickly), we use the 
analytical solution for $k = 0$ as zero-order approximation, and apply WKB to obtain a better 
solution. Finally, we find a solution for the linearized evolution equation of $\varphi$ which is 
proportional to $1/\eta$, thus decreases with time. This could be a disastrous for this 
model, because this leads to a dark energy with $w > -2/3$ which is already ruled out by 
observations. Nonetheless, when the full nonlinear equation is considered, although we cannot 
solve it, there is evidence that under special conditions a roughly constant amplitude - a 
tracking solution - similar to observed dark energy can be obtained. In fact, expectation 
values of type $\langle{\phi}^{i}\rangle$ induce negative power of $\varphi$ into the evolution 
equation of $\varphi$ because $C_k$ is proportional to $\varphi^{-1}$, see equation 
(\ref{condexpg}). As we mentioned in the Introduction, polynomials with negative power are 
proved to have a tracking solution. A counting of power of self-interaction terms after 
replacing expectation values with their approximate solution show that for $n \leq 3$ there 
exits a tracking solution. For $n = 4$ the decrease rate of the condensate density can be 
enough slow to be consistent with present observations. In 4D spacetimes these potentials are 
the only renormalizable self-interactions for a quantum scalar field model. Giving the fact 
that we did not impose any constraint on renormalizability of the model, these results are very 
interesting and encourage more work on quantum description and origin of dark energy. 

If these conclusions are confirmed by a more precise numerical calculations, they would be a 
proof of the reign of quantum mechanics at largest scales in the Universe because it is the 
quantum coherence of dark energy that saves it from being diluted by the expansion. The 
dominance of dark energy at late times in one hand proves that in contrast of general believes, 
the Universe is dominantly in a coherent quantum state, and in the other hand dark energy 
provides a natural envionment for decoherence of other constituents.

\section {\normalsize Vacuum energy} \label{sec:vacuum}
In QFT energy is calculated as the expectation value of classical expression for the energy 
momentum tensor $T^{\mu\nu}$ in which classical field is replaced by its quantized counterpart:
\bea
E &=& \langle \psi | \int d^4 k \delta (k_0^2 - m^2) T^{00} | \psi \rangle = 
\frac{1}{2} \langle \psi | \int d^3 \omega_k  (a^\dagger_k a_k + a_k a^\dagger_k)| 
\psi \rangle = \nonumber \\
&&\langle \psi | \int d^3 \omega_k  (\hat {N}_k + \frac{1}{2})| \psi \rangle, \quad  
\omega_k \equiv \vec{k}^2 + m^2
\eea
Vacuum energy is defined as $|\psi\rangle = |0\rangle$, thus $E_{vac} = \frac{1}{2}\int d^3 
\omega_k \rightarrow \infty$. To regularize this integral usually a UV cutoff is imposed that 
leads to a finite but very large value for the energy density of vacuum. In QFT in Minkovsky 
space without gravity ordering operator is imposed to the above definition i.e. 
$\langle \psi |:T^{00}:| \psi \rangle$ is used. This simple operation removes the constant 
(infinite) term and makes a strictly zero vacuum energy. When gravity is present, it is usually 
supposed that ordering operator cannot be applied because it shifts the energy, an unauthorized 
operation in the context of general relativity and gravity that define an absolute reference 
for energy. 

The constant term in $\langle \psi |T^{00} | \psi \rangle$ is due to noncommutative creation 
and annihilation operators in field theory. The presence of the constant term looks like the 
memory of spacetime or the ghost of a particle, i.e. when a particle is created, its 
annihilation does not completely restore the initial state. This can be interpreted as a 
manifestation of energy conservation. In fact, considering two operators $a^{\dagger}a$ 
and $aa^{\dagger}$, their physical interpretations are very different. The former counts the 
number of particles in a state. For an asymptotically free system, it behaves as a detector 
of particles without changing their state. By contrast, operator $aa^{\dagger}$ first creates a 
particle i.e. changes the energy of the system by an amount equal to the energy of the 
particle. Because in general relativity energy and momentum are locally conserved, this 
operatation necessarily violates the closeness of the system and must actually play the role 
of a bridge between the system - state - under consideration and another system that provides 
the energy. Moreover, energy and momentum are eigen values of translation 
operator. Therefore, the application of right part of this operator changes the translation 
(symmetry) state of the system. Because symmetry is related to information, the return of 
energy to its initial reservoir i.e. the annihilation of the particle restore the energy state, 
but quantum mechanics tells us that it does not restore the information. This is another 
manifestation of nonlocality or state collapse in quantum mechanics. Nonetheless, if we are 
only interested in energy conservation, annihilation restores energy-momentum state and in 
this regard the system should be considered as unchanged. Base on this argument we suggest that 
operator ordering must be applied to $\langle \psi |T^{00} | \psi \rangle$ even in the context 
of general relativity.

\small
\bibliographystyle{apalike}

\end{document}